\documentclass[11pt]{amsart}

\reversemarginpar
\newcommand{\commentout}[1]{}

\usepackage{amsmath}
\usepackage{amssymb}

\newcommand{\cS}{{\mathcal S}}
\newcommand{\gradx}{\nabla_{\bx}}
\newcommand{\Hbar}{\bar{H}}
\newcommand{\gradp}{\nabla_{\bp}}
\newcommand{\half}{\frac{1}{2}}
\newcommand{\nwc}{\newcommand}
\newcommand{\xvec}{\vec{\mathbf x}}

\newcommand{\lt}{\left}
\newcommand{\rt}{\right}

\newcommand{\bx}{\mathbf x}
\nwc{\xbar}{\bar{\mathbf x}}
\nwc{\pbar}{\bar{\mathbf p}}
\nwc{\zbar}{\bar{z}}
\nwc{\zhat}{\hat{z}}
\nwc{\zm}{{z_m}}
\nwc{\zo}{{z_0}}
\newcommand{\bp}{\mathbf p}
\newcommand{\by}{\mathbf y}

\newcommand{\bq}{\mathbf q}
\newcommand{\bw}{\mathbf w}
\newcommand{\br}{\mathbf r}

\nwc{\nwt}{\newtheorem}

\nwt{prop}{Proposition}
\nwt{proposition}{Proposition}
\nwt{lemma}{Lemma}
\nwt{theorem}{Theorem}
\nwt{cor}{Corollary}
\nwt{remark}{Remark}
\nwt{definition}{Definition} 

\nwc{\bal}{\begin{align}}
\nwc{\be}{\begin{equation}}
\nwc{\ben}{\begin{equation*}}
\nwc{\bea}{\begin{eqnarray}}
\nwc{\beq}{\begin{eqnarray}}
\nwc{\bean}{\begin{eqnarray*}}
\nwc{\beqn}{\begin{eqnarray*}}
\nwc{\beqast}{\begin{eqnarray*}}

\nwc{\eal}{\end{align}}
\nwc{\ee}{\end{equation}}
\nwc{\een}{\end{equation*}}
\nwc{\eea}{\end{eqnarray}}
\nwc{\eeq}{\end{eqnarray}}
\nwc{\eean}{\end{eqnarray*}}
\nwc{\eeqn}{\end{eqnarray*}}
\nwc{\eeqast}{\end{eqnarray*}}

\nwc{\tx}{\tilde{\bx}}
\nwc{\tp}{\tilde{\bp}}
\nwc{\tr}{\tilde{\br}}
\nwc{\tw}{\tilde{\bw}}
\nwc{\ep}{\varepsilon}
\nwc{\ept}{\epsilon}
\nwc{\vrho}{\varrho}
\nwc{\orho}{\bar\varrho}
\nwc{\ou}{\bar u}
\nwc{\vpsi}{\varpsi}
\nwc{\lamb}{\lambda}
\nwc{\wep}{W^\ep}

\nwc{\partz}{\frac{\partial }{\partial z}}
\nwc{\partt}{\frac{\partial }{\partial \tau}}

\nwc{\nn}{\nonumber}
\nwc{\bm}{\boldmath}
\nwc{\mf}{\mathbf}
\nwc{\mb}{\mathbf}
\nwc{\ml}{\mathcal}

\nwc{\bD}{{\mb D}}

\nwc{\IA}{\mathbb{A}} 
\nwc{\IB}{\mathbb{B}}
\nwc{\IC}{\mathbb{C}} 
\nwc{\ID}{\mathbb{D}} 
\nwc{\IM}{\mathbb{M}} 
\nwc{\IP}{\mathbb{P}} 
\nwc{\II}{\mathbb{I}} 
\nwc{\IE}{\mathbb{E}} 
\nwc{\IF}{\mathbb{F}} 
\nwc{\IG}{\mathbb{G}} 
\nwc{\IN}{\mathbb{N}} 
\nwc{\IQ}{\mathbb{Q}} 
\nwc{\IR}{\mathbb{R}} 
\nwc{\IT}{\mathbb{T}} 
\nwc{\IZ}{\mathbb{Z}} 

\nwc{\cE}{{\ml E}}
\nwc{\cP}{{\ml P}}
\nwc{\cL}{{\ml L}}
\nwc{\cN}{{\ml N}}
\nwc{\cU}{{\ml U}}
\nwc{\cR}{{\ml R}}
\nwc{\cV}{{\ml V}}
\nwc{\cW}{{\ml W}}
\nwc{\cT}{{\ml T}}
\nwc{\crV}{{\ml V}_{(\delta,\rho)}}
\nwc{\cC}{{\ml C}}
\nwc{\cA}{{\ml A}}
\nwc{\cK}{{\ml K}}
\nwc{\cB}{{\ml B}}
\nwc{\cD}{{\ml D}}
\nwc{\cF}{{\ml F}}
\nwc{\cM}{{\ml M}}
\nwc{\cG}{{\ml G}}
\nwc{\cH}{{\ml H}}
\nwc{\bk}{{\mb k}}
\nwc{\cQ}{{\ml Q}}
\nwc{\cO}{{\ml O}}
\nwc{\cJ}{{\ml J}}

\nwc{\mint}{{\int\cdot\int}}

\begin{document}

\title{Phase Space Models for 
Stochastic  Nonlinear Parabolic Waves: Wave Spread and
Singularity}

\author{Albert C. Fannjiang}
\thanks{Department of Mathematics,
University of California at Davis,
Davis, CA 95616, 
E-mail: cafannjiang@ucdavis.edu. 
Research partially supported by
the Centennial Fellowship from the American
Mathematical Society,
the UC Davis Chancellor's Fellowship
and  U.S. National Science Foundation grant DMS 0306659.
}

\begin{abstract}
We derive  several kinetic equations 
to model the large scale, low Fresnel number
behavior of the nonlinear Schrodinger (NLS)
equation with a rapidly fluctuating random
potential.  There are three types of kinetic
equations the longitudinal, the transverse
and the longitudinal with friction. 
For these nonlinear kinetic equations we address two
problems: the rate of dispersion and the singularity
formation. 

For the problem of dispersion, we show
that the kinetic equations of the longitudinal type
produce the cubic-in-time law, that the
transverse type produce the
quadratic-in-time  law  and that the
one with friction produces
the linear-in-time law  for
the variance prior to any singularity.

For the problem of singularity, we show
that the singularity and blow-up conditions
in the transverse
case remain the same as those for the homogeneous
NLS equation with critical or supercritical self-focusing
nonlinearity, but they
have changed in
the longitudinal case and in the frictional case
due to the evolution of the Hamiltonian.
\end{abstract}
\maketitle

\section{Introduction}

In this paper we consider nonlinear Schr\"odinger (NLS)
equation with a random potential
\beq
\label{para2}
\label{nls}
i\partz
\Psi(z,\bx)+\frac{\gamma}{2}\Delta_\bx\Psi(z,\bx)
+\gamma^{-1}g|\Psi|^{2\sigma}\Psi(z,\bx)+
\mu V(z L_z,\bx L_x)\Psi(z,\bx)&=&0,\\
\bx\in \IR^d, \quad \sigma >0&&\nn
\eeq
where $\gamma$ is the Fresnel number (defined below),
$\mu$ is the linear coupling coefficient for
the random potential $V$, which is rescaled by
two large parameters $L_z$ and $L_x$,
and 
$g$ is nonlinear coupling coefficient with
$g>0$ representing the self-focusing case
and $g<0$ the self-defocusing case. 
Here $\sigma$ is a positive constant and
$\sigma=1$ corresponds to the {\em cubic}
NLS equation. We are particularly
interested in the regime of low Fresnel number
$\gamma\ll 1$ with a rapidly oscillating
potential $L_x, L_z, \mu \gg 1$.

Finite-time blow-up or wave collapse is a well-known
effect for the self-focusing, (super)-critical ($d\sigma
\geq 2$)
 NLS equation without a random potential \cite{SS}. In
this case the nonlinear focusing effect dominates over
the linear diffraction effect of $i\Delta$. 
The question remains whether the added random
potential and its scattering effect would
 prevent the formation of singularity.
Given the wide range of scales present in
this problem the numerical simulation as well
as theoretical analysis
is undoubtedly extremely challenging.
Although current numerical results, e.g. \cite{DD}, have already
indicated that a white-noise-in-$z$ potential
without the feature of rapid fluctuation in $\bx$
does not prevent blow-up. Indeed,
without
the self-averaging effect of a $\bx$-rapidly
fluctuating potential the large  $z$-fluctuations in
the white-noise potential may drive the system to a
state with low, negative Hamiltonian, thus
developing singularities in very short time. These
singularities in our view are the small scale,
randomly fluctuating singularities which may be
delayed or removed in the presence of a
$\bx$-rapidly fluctuating potential.

To capture the robust, large scale blow-ups we
derive several phase-space model equations
corresponding to different scalings (in
$\gamma,\mu, L_x, L_z$). We  use these kinetic
equations to elucidate the problems of singularity
formation and rate of dispersion. The main
ingredient of our analysis is the 
variance identity for these phase space
transport equations. In general, we
show that random scattering changes the
blow-up conditions for the homogeneous
NLS eq. but does not prevent
singularity formation on the large
scales (see Conclusion for the
discussion and summary of our results).
We also give various upper bounds for
the blow-up time.

As a simple by-product of the variance
identity, we also derive various
bounds on
the rate of dispersion for random 
Schr\"odinger  waves, including exact
expressions in the critical case,
$d\sigma=2$. We show that
depending on the random scattering mechanism, one
can have the cubic ($z^3$), the quadratic ($z^2$)
and the linear ($z$) rate of dispersion prior
to singularity, if there is any.
To our knowledge, these
are significant improvements over
current results (e.g.
\cite{BL}, \cite{Tch},  \cite{EKS}) which
are mostly for the linear problem.

The cubic NLS equation is a model equation in nonlinear
optics, describing the
spatial distribution of the stationary electromagnetic
field in a nonlinear medium  with the Kerr effect 
\cite{NM}, \cite{HK}. Here let us give a brief derivation
of the NLS equation with a random potential.

The electric field  $E$ with a fixed polarization  in a
lossless medium satisfies the Helmholtz equation
\[
\nabla_{\xvec}^2 E(\xvec)+k^2 n^2E(\xvec)=0,\quad
\xvec\in \IR^3
\]
where  $k$ the
wavenumber and ${n}(\xvec, |E|^2)>0 $ is
the refractive
index of the medium.
For a weakly nonlinear  and weakly fluctuating  inhomogeneous
medium we write 
\[
n(\xvec, |E|^2)=\bar{n}(1+\tilde{n}(\xvec))
+ {n_2}|E|^2
\]
with
$\tilde{n}\ll 1
$ and the Kerr coefficient $n_2$ being small. Consider the
approximate equation
\beq
\label{helm}
\nabla^2_{\xvec} E(\xvec)+k^2\bar{n}^2
E(\xvec)=-2\bar{n}^2\tilde{n}(\xvec)
E(\xvec)-2\bar{n}n_2|E|^2E(\xvec).
\eeq

The electric field
is taken as
\[
E(z, \bx)=\Psi(z,\bx)
\exp{(i\bar{n}k z)},\quad
\xvec=(z,\bx)\in
\IR^{3}
\]
where $z$ is the coordinate in the
direction of propagation and
$\bx$ is the  transverse coordinates. The modulation
$\Psi$ is assumed to be slowly varying 
on the scale $k^{-1}$.

Substituting the ansatz into
eq. (\ref{helm})  and making the
parabolic approximation we obtain
for the modulation $\Psi$ the equation
\beq
\label{para}
ik \bar{n}\partz \Psi(z,\bx)+\half\nabla_\bx^2
\Psi(z,\bx) +k^2\bar{n}n_2
|\Psi|^2\Psi(z,\bx)+k^2\bar{n}^2
\tilde{n}(z,\bx)\Psi(z,\bx)=0.
\eeq

Let us write 
\[
\tilde{n}(z,\bx)=\nu V(z,\bx)
\]
where  $\nu\ll 1$ is the standard
deviation of $\tilde{n}$
and $V(z,\bx)$ is a zero-mean, $z$-stationary,
$\bx$-homogeneous random field. Let $L_x, L_z\gg 1$
be the two length scales of the wave beam
in the transverse and longitudinal directions,
respectively and  
\[
\tilde{\bx}=\bx/L_x,\quad
\tilde{z}=z/L_z,
\]
be the rescaled variables.
Let us introduce the dimensionless parameters
\[
\mu=k\bar{n}L_z \nu,\quad
\gamma=\frac{L_z}{k\bar{n}L_x^2},\quad
 g={L_z n_2k \gamma}.
 \]
Writing the wave field $\Psi$ in the new variables $\tilde{\bx},
\tilde{z}$,
dropping the tilde 
and rescaling 
we obtain  (\ref{nls}) with $\sigma=1$.

The critical case of the cubic NLS with $d=2$ and $g>0$
is our primary example; another is the 
sub-critical case of a planar
waveguide (thin film) with $d=1$. The
opposite case of a self-defocusing Kerr
nonlinearity occurs in semiconductor
waveguide \cite{MPL}, \cite{Ma}.
Similar equations also arise in many other contexts
such as the Langmuir wave in plasma and Bose-Einstein condensation,
see \cite{SS} and references therein.
Therefore, we will formulate the results for the
general power-law NLS eq. (\ref{nls}) with transverse
dimension $d\geq 1$. Because the variable $z$ is time-like
we will refer to it as ``time'' occasionally, especially
in the discussion of finite-time singularity.

\section{Wigner distribution and Wigner-Moyal equation
} We consider  several families of
scaling limits, as
$\gamma\to 0, L_z, L_x\to\infty$, which are
first distinguished by whether 
\beq
\label{scale}
\theta\equiv\gamma L_x\to 0\quad
\hbox{or}\quad
\theta=1.
\eeq
This, of course, is not sufficient to ensure
the existence of scaling limit until we specify
the strength of $\mu$. 

Our phase-space model equations for
the low Fresnel number regime is based
on the Wigner equations.
We consider
the Wigner distribution of the form
\beq
\label{pure}
 W(\bx,\bp) =   \frac{1}{(2\pi)^d}
           \int e^{-i \bp \cdot \by}
         \Psi\lt(z,\bx+\frac{\gamma \by}{2}\rt)
		{\Psi^*\lt(z,\bx-\frac{\gamma \by}{2}\rt)}
d\by.  
\eeq
The Wigner distribution 
has many important properties.
For instance,
it is real and its $\bp$-integral is the modulus square of the
wave
function,
\bea\label{2.2.2}
\int _{\IR^{d}}W^\ep(\bx,\bp)d\bp=|\Psi(\bx)|^2\equiv
\rho(\bx),
\eea
so we may think of $W(\bx,\bp)$ as wave number-resolved mass density.
Consequently, the mean $M_x$ and variance $V_x$ of $\bx$
are given by, respectively,  
\beqn
\xbar&=&\int \bx W(\bx,\bp)d\bx d\bp,\\
V_x&=&\int |\bx-\xbar|^2 W(\bx,\bp)d\bx d\bp\\
&=&S_x-\xbar^2
\eeqn
where $S_x$ is the second moment of $\bx$
\[
S_x=\int |\bx|^2 W(\bx,\bp)d\bx d\bp.
\]
Also, we have that
\beqn
\int_{\IR^d}W^\ep(\bx,\bp)d\bx&=&(\frac{2\pi}{\gamma})^{d}
|\widehat\Psi^\ep|^2(\bp/\gamma).
\eeqn
and the energy flux
\bea\label{2.2.3}
\frac 1{2i}(\Psi\nabla\Psi^*-\Psi^*\nabla\Psi)=
\int_{\IR^d}\bp W^\ep(\bx,\bp)d\bp
\eea
so that the mean and variance  of $\bp$,
$V_p$, are given by, respectively
\bean\label{2.2.4}
\pbar&=&\int \bp W(\bx,\bp)d\bx d\bp,\\
V_p&=&\int
|\bp-\pbar|^2W(\bx,\bp)d\bx d\bp
=S_p-\pbar^2
\eean
where $S_p$ is the second moment of $\bp$
\[
S_p=\int |\bp|^2 W(\bx,\bp)d\bx d\bp.
\]

In view of these properties it is
tempting to think of the Wigner distribution as
a phase-space probability density, which is unfortunately not
the case, since it is not everywhere non-negative
(it is always real-valued though).

It is straightforward to derive the
 closed-form equation for the Wigner distribution
 \cite{rad-arma}
\beq
\label{nwm'}
\frac{\partial W}{\partial z}
+{\bp}\cdot\gradx W+\cU_\gamma W
+\mu \cV W=0,\label{wig}
\eeq
with the Moyal operators
\beq
\label{cu}
\cU_\gamma W( \bx,\bp)&=&
i\int e^{i\bq\cdot\bx}
\gamma^{-1}\lt[W(\bx,\bp+{\gamma}\bq/2)-W(\bx,\bp-
{\gamma}\bq/2)\rt]
\widehat{U}(z,\bq)d\bq,\quad U= g|\Psi|^{2\sigma}\\
\cV W(z, \bx,\bp)&=&
i\mu\int e^{i\bq\cdot\bx L_x}
\lt[W_z(\bx,\bp+{\theta}\bq/2)-W_z(\bx,\bp-
{\theta}\bq/2)\rt]
\widehat{V}({z}L_z,d\bq)
\label{cv}
\eeq
where 
$\hat{U}, \hat{V}$ denote
the Fourier transforms of $U,V$, respectively.
Formally we see that  as
$\gamma\to 0$
\beq
\label{go}
\cU_\gamma W(\bx,\bp)\to \cU_0W(\bx,\bp)\equiv \gradx
U(z,\bx)\cdot\nabla_\bp W(\bx,\bp).
\eeq
We use
the following definition of the Fourier transform 
and inversion:
\beqn
\cF f(\bp)&=&\frac{1}{(2\pi)^d}\int e^{-i\bx\cdot\bp}
f(\bx)d\bx\\
\cF^{-1} g(\bx)&=&\int e^{i\bp\cdot\bx} g(\bp) d\bp.
\eeqn
We shall refer to eq. (\ref{wig})
as the nonlinear Wigner-Moyal (NWM) equation. 
We only need to  consider the weak formulation
\beq
\partz \int \Theta Wd\bx d\bp
-\int \bp\cdot\gradx \Theta Wd\bx d\bp
-\int \cU_\gamma \Theta Wd\bx d\bp
-\mu \int \cV \Theta W d\bx d\bp
=0
\label{nwm}
\eeq
for smooth, rapidly decaying test functions
$\Theta$.

One advantage of working with eq. (\ref{wig})
is that one can use it to evolve
the mixed-state initial condition,
instead of the pure-state one given
in (\ref{wig}). This is important in the
context of modeling quantum open systems.
A mixed state Wigner distribution is
a convex  combination of the pure-state
Wigner distributions (\ref{pure}) described as follows.

Let $\{\Psi_\alpha\}$ be a family of  $L^2$ functions
parametrized by $\alpha$ which is weighted by
a probability measure $P(d\alpha)$ .
 Denote the pure-state Wigner distribution
(\ref{wig}) by $W[\Psi]$. A mixed-state Wigner
distribution is given by
\beq
\label{mixed}
\int W[\Psi_\alpha]P(d\alpha).
\eeq
The limits as $\gamma\to 0$ of the mixed state
Wigner distributions constitute the so called
Wigner measures  which are always positive
\cite{GL}, \cite{GMMP}, \cite{LP}. Evolution by eq.
(\ref{nwm}) preserves the form (\ref{mixed}). In
particular, for such initial data
we have
\beq
\label{pos1}
\int W(z,\bx,\bp)d\bp&\geq& 0,\quad \forall \bx \in
\IR^d\\
\int W(z,\bx,\bp)d\bx &\geq& 0,\quad \forall \bp\in
\IR^d
\label{pos2}
\eeq

Multiplying (\ref{nwm}) by $W$ and integrating
by parts we also see that the evolution preserves
the $L^2(\IR^{2d})$-norm of $W$, i.e.
\[
\partz \int | W|^2 d\bx d\bp =0.
\]

Let us be more explicit about the
random potential $V(z,\bx)$. We
assume that $z$-stationary, $\bx$-homogeneous random
field $V$ is square-integrable and admits the spectral
representation
\[
V(z,\bx)=\int \exp{(i\bp\cdot\bx)}\hat{V}(z,d
\bp)
\]
with the  $z$-stationary spectral measure
$\hat{V}(z,\cdot)$ satisfying
\[
\IE[\hat{V}(z,d\bp)\hat{V}(z,d\bq)]
=\delta(\bp+\bq)\Phi_0(\bp)d\bp d\bq. 
\]
Here $\IE$ denotes the ensemble expectation 
and the function $\Phi_0(\bp)$ is the
transverse power spectrum density which is
assumed to be rapidly decaying
as $|\bp| \to 0$ or $|\bp|\to\infty$.
The transverse power spectrum  density is
related to the full power spectrum density
$\Phi(w,\bp)$ in the following way
\[
\Phi_0(\bp)=\int \Phi(w,\bp)dw
\]
with $\Phi(\bk)$
satisfying $\Phi(\bk)=\Phi(-\bk),\forall
 \bk=(w,\bp)\in \IR^{d+1}.$
Without loss of generality we will also assume
\beq
\label{sym}
\Phi(w,\bp)=\Phi(-w,\bp)=\Phi(w,-\bp)
=\Phi(-w,-\bp).
\eeq

\section{Phase space transport equations}

\subsection{Linear kinetic equations.} First we
summarize  what has been established in the linear case
when $U(z,\bx)$ is a {\em given} function. In this case
we refer to eq. (\ref{wig}) with a given $U$ 
as the linear Wigner-Moyal (LWM) equation.
The principal feature of the scaling 
is the separation of scales in the given
potential $U$ and the random potential $V$ as 
they
appear in the LWM equation, c.f. (\ref{cu})
and (\ref{cv}).
Under pretty general conditions (the most important of
which being the integrability of the maximum correlation
coefficient of $V(z,\cdot)$ as $\bx$-homogeneous-field-valued
$z$-stationary process)
one can prove that, as $\gamma\to 0$ and $ \mu, L_x,
L_z\to\infty$, the weak solution of the LWN
equation converges in law to the weak solution of
the linear Boltzmann (LB) equation or the linear
Fokker-Planck (LFP) equation, described below,
depending on whether
$\theta$ also tends to zero or not
 \cite{rad-arma},\cite{saq} (see also 
\cite{BPR}).

\bigskip

\noindent {\bf  Linear Boltzmann
equation ($\theta=1$): }

\beq
\label{lb}
\lefteqn{\partz W(z,\bx,\bp)+\bp\cdot\gradx W
(z,\bx,\bp)+\gradx U(z,\bx)\cdot\nabla_\bp W(z,\bx,\bp)}\\
&=&2\pi \int K(\bp,\bq) [W(z,\bx,\bq)-
W(z,\bx,\bp)]d\bq\nn
\eeq
with a nonnegative kernel $K(\bp,\bq)$ given
respectively as follows.

\begin{description}
\item[(a)] If $\mu\sim \sqrt{L_z}, L_x\ll
L_z$ then
\beq
\label{K1}
K(\bp,\bq)
=\Phi(0,\bq-\bp)
\eeq
\item[(b)] If $\mu\sim \sqrt{L_x}, L_z\ll L_x\ll L_z^{4/3}, 
d\geq 3$
then 
\beq
\label{K2}
K(\bp,\bq)
=\delta(\frac{|\bq|^2-|\bp|^2}{2})
\lt[\int \Phi(w,\bq-\bp)dw\rt].
\eeq
\item[(c)] If $\mu\sim \sqrt{L_z}, L_x\sim
L_z$ then
\beq
\label{K3}
K
(\bp,\bq)
=
\Phi(\frac{|\bq|^2-|\bp|^2}{2},\bq-\bp).
\eeq

\end{description}

\bigskip

\noindent {\bf  Linear
Fokker-Planck equation ($\theta\to 0$):}
\beq
\label{lfp}
\lefteqn{\partz W(z,\bx,\bp)+\bp\cdot\nabla_\bx W
(z,\bx,\bp)+\nabla_\bx U(z,\bx)\cdot\nabla_\bp
W(z,\bx,\bp)}\\ &=&\nabla_\bp \cdot \bD\nabla_\bp
W(z,\bx,\bp)\hspace{5cm}\nn
\eeq
with a symmetric,
nonnegative-definite matrix $\bD$
given as follows.
\begin{description}
\item[(a)] If $\mu\sim \theta^{-1}
\sqrt{L_z}, 
L_x\ll L_z$ then
\beq
\label{D1}
\bD=\pi\int \Phi(0,\bq)\bq\otimes\bq d\bq.
\eeq
\item[(b)] If $\mu\sim \theta^{-1}
\sqrt{L_x}, L_z\ll L_x\ll  L_z^{4/3}, d\geq 3$ then
\beq
\label{D2}
\bD(\bp)=\pi |\bp|^{-1}
\int_{\bp\cdot\bp_\perp=0} \int
\Phi(w,\bp_\perp) d w\,\,
\bp_\perp\otimes\bp_\perp
d\bp_\perp.
\eeq
\item[(c)] If $\mu\sim \theta^{-1}
\sqrt{L_z},  L_x\sim L_z$ then
\beq
\label{D3}
\bD(\bp)=\pi\int \Phi(\bp\cdot\bq,\bq)
\bq\otimes\bq d\bq.
\eeq
\end{description}
We shall use $\cL$ to denote 
either the scattering operator on
the right side of eq. (\ref{lb}) or
the diffusion operator on the right side
of eq. (\ref{lfp}).  The self-adjoint operator $\cL$ is
non-positive definite and represents various
decoherence effects due to random fluctuations of the medium,
see (\ref{3.3}). The evolution by Eq. (\ref{lb}) or
(\ref{lfp}) preserves the mixed-state structure
(\ref{mixed}) and hence the positivity
(\ref{pos1}), (\ref{pos2}).
 
 We shall
refer to regime (a), for either (\ref{lb})
or (\ref{lfp}), as the {\em
longitudinal} case and regime (b)
as the {\em transverse} case  because the longitudinal
and the transverse scales are dominant, respectively.
Regime (c) is the borderline case.  We will
consider hereafter the longitudinal and 
transverse cases only.

We note that the additional restrictions of $L_x\ll 4/3
L_z$ and $d\geq 3$ in regime (b) are due to technical
reasons.
 We believe that
the results should hold for any $L_z\ll L_x $, 
$d\geq 2$ along with the  other assumptions used in
proving the above scaling limits (see \cite{rad-arma},
\cite{saq} for details). The mean field result for a
Gaussian potential with 
$L_z=0$ and $d\geq 3$ has been established
previously \cite{Sp}, \cite{EY} (see also \cite{PV}).

\subsection{Nonlinear kinetic equation.} When
$U=g|\Psi|^{2\sigma}$, the convergence of the above
scaling limits is not known. But because the phase of
$\Psi$ is canceled  in the expression of $U$ one can
reasonably expect that, at least
before any singularity formation, $U$ has significantly
less oscillation than $\Psi$ and $V(zL_z, \bx L_x)$
as appearing in the NLS equation (\ref{nls}).
The property of {\em separation of scales}
in $U$ and $\Psi$ holds in the linear
case and will be assumed to hold
prior to singularity formation.

Hence we will assume  below
the validity of  the above scaling limits
for the nonlinear case and use the 
kinetic equation (\ref{lb}) or (\ref{lfp})
with $U=g|\Psi|^{2\sigma}$
to investigate the issue of random diffraction and
 singularity formation
in the case of self-focusing ($g>0$), (super)-critical
($d\sigma\geq 2$) nonlinearity as well
as the question of rate of spread in all situations.
Such a model with the kernel (\ref{K2})
was considered in \cite{nls}.
The Fokker-Planck equation with the diffusion 
coefficient (\ref{D1}) is an 
example of those which describe open quantum
systems in contact
with a heat bath of linear oscillators in
thermal equilibrium (see, e.g., \cite{CL}, \cite{Dio},
\cite{AM}, \cite{VK}). In this context, $z$ would be the
physical time, $\bx$ the physical coordinates and
$\bp$ the momentum of the quantum particle.
We will discuss the master equation for the open system
including dissipation mechanism in the final section.

Let us state the nonlinear kinetic equation
which we will analyze subsequently:
\beq
\label{qt}
\partz W(z,\bx,\bp)+\bp\cdot\nabla_\bx W
(z,\bx,\bp)+\cU_0
W(z,\bx,\bp)=\cL W(z,\bx,\bp)
\eeq
where $\cU_0$ is given by  (\ref{go})
and $\cL$ is either the linear
scattering operator
\[
\cL W=2\pi \int K(\bp,\bq) [W(z,\bx,\bq)-
W(z,\bx,\bp)]d\bq
\]
or the linear diffusion operator
\[
\cL W= \nabla_\bp \cdot \bD\nabla_\bp
W(z,\bx,\bp).
\]
The weak formulation of (\ref{qt})
is  given by 
\beq
\label{qtg}
\partz \int \Theta Wd\bx d\bp
-\int \bp\cdot\gradx \Theta Wd\bx d\bp
-\int \gradx U\cdot\gradp\Theta Wd\bx d\bp
-\int \cL \Theta W d\bx d\bp
=0
\eeq
for smooth, rapidly decaying test functions
$\Theta$.
The evolution preserves the positivity of
the initial condition which is consistent
with the fact that the Wigner measures
are positive and will be assumed
subsequently (see the comment following
(\ref{mixed})).

The nonlinear kinetic eq. (\ref{qtg}) preserves
the total mass, i.e.
\[
\partz N=0,\quad N= \int W(z,\bx,\bp)d\bx d\bp
\]
but in general decreases the $L^2$-norm
\beq
\label{3.3}
\partz \int |W|^2(z,\bx,\bp)d\bx d\bp
=\int W \cL  W d\bx d\bp\leq 0
\eeq
because the operator $\cL$ is non-positive definite.
The inequality (\ref{3.3}) expresses certain
irreversibility as a result of the weak convergence 
of solutions \cite{rad-arma}. One
can absorb the effect of the total mass $N$ into
$g$ by the obvious rescaling of $W$ in the eq.
(\ref{qtg}). Henceforth we assume
that
\[
N=1
\]
which is the case when in (\ref{mixed})
\[
\int |\Psi_\alpha(\bx)|^2d\bx=1
\]
for $P$-almost all $\alpha$.
\commentout{
In this circumstance, the $L^2$-norm of
$W$ can be expressed as
\[
\int |W|^2d\bx d\bp=\sum _{j=1}^\infty p_j^2\leq 1.
\]
}

\subsection{Local existence}
A natural space of initial data and solutions is
the space $\cS$ of the non-negative measures with
square integrable density $W$
\[
\int |W|^2d\bx d\by<\infty, 
\]
finite Dirichlet form
\[
-\int W\cL Wd\bx d\bp<\infty
\]
and finite, positive variances
\[
V_x=\int |\bx-\xbar|^2 Wd\bx d\bp=S_x-|\xbar|^2\in
(0,\infty),
\quad V_p=\int |\bp-\pbar|^2 Wd\bx d\bp
=S_p-|\pbar|^2\in (0,\infty)
\]
where $S_x, S_p$ are the second moments
\beq
\label{sec}
S_x=\int |\bx|^2 Wd\bx d\bp\in (0,\infty),\quad
S_p=\int |\bp|^2Wd\bx d\bp\in (0,\infty).
\eeq
In addition, we shall also assume that
the initial data have a finite
 Hamiltonian  $H$ 
\[
H=\frac{1}{2}S_p-\frac{g}{\sigma+1}
\int \rho^{\sigma+1}d\bx\in (-\infty,\infty), \quad
\rho(\bx)=\int Wd\bp
\] 
the first term of which is the kinetic energy
and the last two terms are the potential energy.
A finite Hamiltonian and a finite variance $V_p$
then imply
\beq
\label{pot}
\int \rho^{\sigma+1}d\bx<\infty,
\eeq
namely  a finite potential energy.

We recall that for the NLS eq. without
a random potential and the scaling (\ref{scale}) the local
existence in  the analogous space of functions
(i.e. $\Psi, \bx \Psi, \nabla \Psi, |\Psi|^{\sigma+1}\in L^2(\IR^d)$)
holds under the additional assumption
\[
0\leq \sigma< \frac{2}{d-2},\quad\hbox{for}\,\,d\geq 3
\]
and $
\sigma\in [0,\infty),\hbox{for}\,\, d\leq 2$,
see \cite{SS} and the references therein. 

The local existence for the nonlinear kinetic
equation (\ref{qtg}) in the space
$\cS$ is assumed throughout the paper and will be
addressed elsewhere.

\subsection{Energy law}
Next, we discuss
the evolution of the Hamiltonian $H$ and the variance
$V_x$. Let us first note the result of $\cU_\gamma$ 
applied to the quadratic polynomials. 
\beq
\label{ga1}
\cU_\gamma \bx&=&0\\
\cU_\gamma \bp&=&i\int e^{i\bq \cdot \bx}\bq
\hat{U}(d\bq)=\gradx U\\
\cU_\gamma |\bx|^2&=&0\\
\cU_\gamma \bx\cdot\bp&=&i\int e^{i\bq\cdot\bx}\bx\cdot\bq
\hat{U}(d\bq)=\bx\cdot\gradx U\\
\cU_\gamma |\bp|^2&=& i\int e^{i\bq\cdot\bx}
2\bp\cdot\bq \hat{U}(d\bq)=2\bp\cdot\gradx U.
\label{ga2}
\eeq
It is noteworthy that the results of the calculation
are 
independent of $\gamma\geq 0$ and identical to
those for $\gamma=0$ (see more on this
in Conclusion).

Consider the mean dynamics for
\beq
\label{1st}
\xbar=\int \bx Wd\bx d\bp,\quad
\pbar=\int \bp Wd\bx d\bp
\eeq
with the mean Hamiltonian defined as
\beq
\label{Hbar}
\Hbar=\frac{1}{2}|\pbar|^2.
\eeq
Using the above and integrating by parts we obtain
the following
\beqn
\partz \pbar&=&\int \cL^*\bp Wd\bx d\bp\\
\partz V_p&=&\int \nabla_\bx U \cdot\bp Wd\bx d\bp
+\int \cL^*|\bp|^2 Wd\bp d\bx\\
\partz \frac{g}{\sigma+1}\int \rho^{\sigma+1}d\bx&=&
\int \nabla_\bx U\cdot \bp Wd \bp d\bx
\eeqn
and hence
\beq
\label{h1}
\partz \Hbar&=& \pbar\cdot\int \cL^*\bp Wd\bx d\bp\\
\partz H&=&\int \cL^*|\bp|^2 Wd\bp d\bx
\label{h2}
\eeq
which will take a more explicit
form once we calculate $ \cL^*\bp, \cL^*
|\bp|^2$ with $\cL^*$ of each case.

As is the case without a random
potential and the scaling limit (\ref{scale})
\cite{SS},
we assume subsequently that
the laws (\ref{h1}), (\ref{h2})
for the Hamiltonian holds for
the local solutions of eq. (\ref{qtg}) in
the space $\cS$.

In the following sections, we first derive the variance
identity for eq. (\ref{qtg}) in the
longitudinal case (regime (a)), then the transverse
case (regime (b)) and finally the case with friction.
As is the case without a random potential and the
scaling limit (\ref{scale}) \cite{SS}, we assume throughout the
paper that the variance identity derived below holds
true for the maximally extended local solutions of eq. (\ref{qtg})
in the space $\cS$.
Let $[0, z_*)$ be the interval on
which the maximally extended local solution
is defined. When $z_*=\infty$ then
the local solutions become  global
solutions; when $z_*<\infty$ then
the solutions are said to develop finite
time singularity.
Our main goal then is to analyze what
these formally derived relations
tell us about singularity formation
and rate of spread of low Fresnel number
nonlinear parabolic waves in random
media.

Before concluding this section let us state  some
elementary inequalities which will be useful later.

An application of the Cauchy-Schwartz inequality
and the marginal positivity (\ref{pos1}), (\ref{pos2})
to the first moments $\xbar, \pbar$
then leads to
\beq
\label{32}
|\xbar|^2\leq S_x,\quad |\pbar|^2 \leq S_p.
\eeq
 Furthermore,
by using the mixed-state structure (\ref{mixed})
in estimating the cross moment
\beq
S_{xp}=\int \bx\cdot\bp Wd\bx d\bp.
\eeq
one deduces that
\beq
\label{cs2}
S^2_{xp}\leq S_x S_p<\infty.
\eeq
Likewise the covariance
\[
V_{xp}=\int (\bx-\xbar)\cdot(\bp-\pbar)Wd\bx d\bp=
S_{xp}-\xbar\cdot\pbar
\]
 can be bounded by $V_x$ and $V_p$ as
\beq
\label{cs}
V_{xp}^2\leq V_x V_p.
\eeq

\section{The longitudinal case}
\subsection{Variance identity}
The variance identity has been long used to
derive the wave collapse condition for 
the NLS without the random potential \cite{SS}.
Below we reformulate it and extend it to
the phase-space models (\ref{qtg}).

We have the following simple calculations for
$\cL^*=\cL$:
\beqn
\cL \bx&=& 0\\
\cL\bp&=&2\pi\int \Phi(0,\bp-\bq)(\bp-\bq)d\bq=0\quad
(\hbox{by (\ref{sym})})\\
\cL |\bx|^2&=&0\\
\cL \bx\cdot\bp&=&
\bx\cdot 2\pi\int \Phi(0,\bp-\bq)(\bp-\bq)d\bq=0\quad
(\hbox{by (\ref{sym})})\\
\cL |\bp|^2&=&-2\pi\int \Phi(0,\bp-\bq)
\lt[ |\bp|^2-|\bq|^2\rt] d\bq \\
&=& -2\pi\int \Phi(0,\bq)(2\bp-\bq)\cdot\bq
d\bq\\
&=& 2\pi\int \Phi(0,\bq)|\bq|^2d\bq\equiv  R.
\eeqn
We have used the linear Boltzmann operator $\cL$
in the above calculation; the same result holds
for the linear diffusion operator $\cL$
for which case 
\[
R=2\times\hbox{trace}(\bD)
\]
where the diffusion matrix $\bD$ is given
by (\ref{D1}).
We shall use the above identities
to perform integrating by parts in
the derivation of the variance identity.

The evolution of the mean position $\xbar$ and
momentum $\pbar$
     is then given by
\beqn
\partz \xbar
&=&\pbar\\
\partz \pbar
&=&\frac{g\sigma}{\sigma+1}
\int \gradx \rho^{\sigma+1}d\bx\\
&=&0,
\eeqn
as a consequence the mean Hamiltonian  
$\Hbar$
is invariant. Moreover, the evolution
of $(\xbar,\pbar)$ for the autonomous  Hamiltonian
system  of the harmonic oscillator.

The evolution of the variance
$V_x$ is given by
\beqn
\partz V_x
&=&2 V_{xp}.
\eeqn

Differentiating $S_{xp}$ we obtain
\beqn
\partz S_{xp}
\nn&=&S_p-\frac{gd\sigma}{\sigma+1}\int 
\rho^{\sigma+1}d\bx d\bp
\eeqn
Hence the second derivative of $V_x$ becomes
\beq
\frac{\partial ^2}{\partial z^2} V_x
&=&4(H-\Hbar)+\frac{2(2-d\sigma)g}{\sigma+1}\int
\rho^{\sigma+1}d\bx.
\label{var1}
\eeq

An alternative expression for the variance identity
is
\beq
\label{var2}
\frac{\partial ^2}{\partial z^2} V_x
 &=&2d\sigma
(H-\Hbar)+(2-d\sigma)V_p.
\eeq
Both forms (\ref{var1}) and (\ref{var2}) will
be used to obtain  dispersion estimates below.

\subsection{Dispersion rate}

Although the medium is lossless, reflected in
the fact that the total mass $N=1$ is conserved,
but the Hamiltonian is not conserved by the evolution
since the scattering with the random potential is
not elastic.
Indeed, its rate of change  is
\beq
\partz H&=& R,\quad H(z)=H(0)+Rz
\eeq
 due to the diffusion-like spread in the momentum
$\bp$.

In the critical case $d\sigma=2$, we have
the exact result
\[
\frac{\partial^2}{\partial z^2}
V_x=4 H-2|\pbar|^2=4(H(0)-\Hbar)+4R z
\]
before any singularity formation
and
hence the following.
\begin{prop} 
If $d\sigma=2$ or $g=0$, then
\beq
\label{exact1}
V_x(z)=V_x(0)+2V_{xp}(0)z+2(H(0)-\Hbar)z^2+
\frac{2R}{3} z^3,\quad z\in [0, z_*).
\eeq

\end{prop}
The analogous result ($V_x\sim z^3$) for the  {\em
linear } Schr\"odinger equation ($d=1, g=0$)
 with a random potential has been proved
previously \cite{BL}, \cite{EKS}.

We have from (\ref{var1})
that
\beq\label{var11}
\frac{\partial ^2}{\partial z^2}
V_x&=&4 H+\frac{(4-2d\sigma)g}{\sigma+1}
\int \rho^{\sigma+1}d\bx-2|\pbar|^2
\eeq
and hence 
\beqn
\frac{\partial ^2}{\partial z^2}
V_x&\leq &4
H-4\Hbar=4(H(0)+Rz)-4\Hbar,\quad\hbox{for}\,\,g(2-d\sigma
)<0\\
\frac{\partial ^2}{\partial z^2}
V_x&\geq &4
H-4\Hbar=4(H(0)+Rz)-4\Hbar,\quad\hbox{for}\,\,g(2-
d\sigma)\geq
0.
\eeqn
On the other hand, from (\ref{var2})
we obtain 
for any $g$
\beqn
\frac{\partial ^2}{\partial z^2} V_x
&\leq&2d\sigma (H-\Hbar),\quad
\hbox{for}\,\, 2-d\sigma\leq 0\\
\frac{\partial ^2}{\partial z^2} V_x
&\geq&2d\sigma (H-\Hbar),\quad
\hbox{for}\,\, 2-d\sigma\geq 0.
\eeqn

Integrating the above inequalities twice, we obtain
the following.
\begin{prop}
The estimates hold
\beqn
V_x(z)&\leq& V_x(0)+2V_{xp}(0)z+2(H(0)-\Hbar)z^2
+\frac{2}{3}Rz^3,\quad\hbox{for}\,\,g(2-d\sigma)\leq
0\\
V_x(z)&\geq& V_x(0)+2V_{xp}(0)z+2(H(0)-\Hbar)z^2
+\frac{2}{3}Rz^3,\quad\hbox{for}\,\,g(2-d\sigma)\geq
0
\eeqn
and
\beqn
V_x(z)&\leq& V_x(0)+2V_{xp}(0)z+d\sigma (H(0)-\Hbar)z^2
+\frac{d\sigma}{3}Rz^3,\quad\hbox{for}\,\,2\leq d\sigma\\
V_x(z)&\geq& V_x(0)+2V_{xp}(0)z+d\sigma (H(0)-\Hbar)z^2
+\frac{d\sigma}{3}Rz^3,\quad\hbox{for}\,\,2\geq d\sigma
\eeqn
for all $z\in [0, z_*)$.

\end{prop}

Therefore
\begin{cor}
Assume $ g< 0$ (hence $H\geq 0$). Then
\beqn
 V_x(z)\leq
V_x(0)+2V_{xp}(0)z+(d\sigma\vee 2) [H(0)-\Hbar]z^2
+\frac{d\sigma \vee2}{3}Rz^3
\eeqn
and
\beq
\label{defocus1}
 V_x(z)\geq
V_x(0)+2V_{xp}(0)z+(d\sigma\wedge 2) [H(0)-\Hbar]z^2
+\frac{d\sigma\wedge 2}{3}Rz^3
\eeq
for $z\in [0, z_*)$
\end{cor}

\subsection{Singularity formation}
\label{sing-sec1}
Finite-time singularity for the
critical or supercritical NLS ($d\sigma \geq 2$)
without a random potential is a well known effect
\cite{SS}. In this case the singularity is the
blow-up type $V_p, \|\rho\|_{\sigma+1}
\to \infty$.  Here we take
(\ref{qtg}) as a model equation to gain some insight
into
singularity formation in
the presence of a random potential.

First we consider the self-focusing case $g\geq 0$.
For
$d\sigma
\geq 2$  one can bound  
$ V_x$ as
\beq
\label{vin}
\label{vin1}
V_x(z)\leq V_x(0)+2V_{xp}(0)z+d\sigma (H(0)-\Hbar) z^2
+\frac{d\sigma R}{3} z^3\equiv F(z)
\eeq
and looks for the situation when
$F(z)$ vanishes.

The sufficient conditions for $F(z)$ to vanish
at a finite positive $z$ are that $F(z)$ takes
a non-positive value  $F(\zo)\leq 0$ at its
 local minimum point $\zo>0$.
The local
minimum point
$\zo$  is given by
\beq
\label{zo}
\zo=\frac{\Hbar-H(0)+
\sqrt{(H(0)-\Hbar)^2-
2RV_{xp}(0)/(d\sigma)}}{ R}.
\eeq

Therefore we are led to the singularity conditions
for $g\geq 0$.
\begin{prop}
\label{sing-prop1} 
For
$d\sigma
\geq 2,g> 0$, the local solution of the nonlinear kinetic
eq. (\ref{qtg})  with the longitudinal randomness 
develops singularity  at a finite time $z_*\leq \zo$
given by (\ref{zo}) under the condition $F(\zo)\leq
0$ and either one of the following conditions
\beq
\label{in}V_{xp}(0)&<&0\\
\label{out} V_{xp}(0) &>&0,\quad 
H(0)<\Hbar-\sqrt{\frac{2R V_{xp}(0)}{d\sigma}}.
\eeq
\end{prop}
\begin{remark}
Clearly, the condition $F(\zo)\leq 0$ requires
$H(0)$ to be
sufficiently below $\Hbar$ by allowing
the potential energy 
\[
-\frac{g}{\sigma+1}\int \rho^{\sigma+1}d\bx d\bp
\]
at $z=0$
to be sufficiently negative.
We will give an explicit blow-up condition for
the supercritical nonlinearity in the next
section.

In the linear or self-defocusing case $g\leq 0$ the
right side of (\ref{exact1}) or (\ref{defocus1}) is
always positive in view of the inequality
\beqn
2|V_{xp}|\leq V_x+V_p,
\eeqn
cf. (\ref{cs}).

\end{remark}
Therefore, the
above result suggests that the scattering term does not
prevent large scale singularity, even though it
appears (from the upper bound $z_0$) that
it takes longer time than the time to singularity
in the homogeneous NLS \cite{SS}.

\commentout{***************************************
**************************************************
Because of eq. (\ref{exact1}) is exact, 
the above singularity condition is also necessary
in the critical case $d\sigma=2$. Moreover
we have
\begin{prop}
Assume $d\sigma=2$ and the wave collapse at one point. The solution of (\ref{qtg})
develops singularity at the finite time
$z_*
$, which equals the middle root of the cubic polynomial (\ref{exact1}),
if and only if $F(\zo)\leq 0$ and either condition (\ref{in}) or
(\ref{out}) holds. 

In case of singularity formation, if $F(\zo)<0$, then
the wave collapses in the following way as $z\to z_*$
\beq
\label{rate}
V_x(z) \sim (z_*-z), \quad z< z_*;
\eeq
if $F(\zo)=0$, then
\beq
\label{rate2}
V_x(z) \sim (z_*-z)^2, \quad z<z_*.
\eeq
\end{prop}
We note that the rate of collapse (\ref{rate}) is
slower than that of the homogeneous critical NLS eq.
which is the same as (\ref{rate2}), see
\cite{SS}.
************************************}

\commentout{*******
Suppose that under the singularity condition
(\ref{in}) or (\ref{out}) $V_x(z)$ reaches
a maximum $\bar{V}_x$ at time $\zbar$ and $V'_x(z)$
becomes negative afterward prior to the singularity
formation. This can be achieved, for instance,
by choosing the parameters so that
$H(0)+R\zo<0$ where $\zo$ given by (\ref{zo})
is the upper bound for $z_*$.
*********}

With the assumption that the variance
identity and the energy laws (\ref{h1}),
(\ref{h2}) hold for the maximally extended local
solutions of eq. (\ref{qtg}) then it is clear that the
singularity is the blow-up type
 \[
 \lim_{z\to z_*}\int\rho^{\sigma+1}d\bx=\infty
 \]
 or equivalently
 \[
 \lim_{z\to z_*}S_p=\infty
\]
by the finitude of the Hamiltonian.

Next we will follow the argument of
\cite{Gl} to show  more explicitly the
blow-up mechanism  in the case  with
the supercritical, self-focusing
nonlinearity  and give a sharper bound
on $z_*$ under certain circumstances. 

\begin{prop}
Suppose 
$d\sigma>2, g> 0$. Then under the
conditions
(\ref{in}),
\beq
\label{blow1}
H(\zo)=H(0)+R\zo\leq 0
\eeq
with $z_0$ given by (\ref{zo}),
$V_x'$ as well as $V_p$ blow up at a finite time
$z_*$ where 
\[
z_*\leq z_0\wedge\frac{2V_x(0)}{V_{xp}(0)(2-d\sigma
)}.
\]
\end{prop}

Let us give the argument below.
Since blow-up is a local phenomenon, $V_x$ is a poor
indicator of  its occurrence. 
To this end a more useful object to consider  is $V_p$.

From (\ref{var2}) it follows
that
\beq
\label{var2.1}
\frac{\partial ^2}{\partial z^2} V_x
&\leq &(2-d\sigma)V_p<0,\quad z< z_*.
\eeq
Hence $V'_x(z)$ is a negative, decreasing function
for $z<z_*$. Also
by the Cauchy-Schwartz inequality, we get
\beq
\label{4.4}
0\leq V^2_{xp}\leq
V_xV_p\leq {V}_x(0)V_p,\quad  0<z< z_*
\eeq
and hence
\beq
\label{47}
V_p\geq \frac{V^2_{xp}}{V_x(0)}.
\eeq

Let $A(z)=-V'_x(z)>0, z<z_*$. We have from
(\ref{var2}) and (\ref{4.4}) the differential
inequality
\beq
\label{grow}
\partz A\geq CA^2,\quad C=\frac{d\sigma-2}{4{V}_x(0)}>0
\eeq
which yields the estimate
\[
A(z)\geq \frac{A(0)}{1-zCA(0)},\quad
z< \frac{1}{CA(0)}
\]
and thus the blow-up of
$-V_x'(z)$ at a finite time.
This along with (\ref{47}) then (\ref{blow1})
implies the divergence of $V_p$ at a finite time
\[
\lim_{z\to z_*}V_p(z)=\infty
\]
with
\beq
\label{4.8}
z_*\leq
\frac{1}{CA(0)}=\frac{2V_x(0)}{V_{xp}(0)(2-d\sigma )}.
\eeq
For high power $d\sigma\gg 2$, (\ref{4.8})
is a better upper bound for $z_*$ than $\zo$
given by (\ref{zo}).

The preceding argument 
demonstrates clearly the blow-up mechanism, namely
the quadratic growth property (\ref{grow}) as well
as gives a sharper bound on the blow-up time
for large $d\sigma$.

\section{The transverse case}
The scattering operator $\cL$ in
this case corresponds to elastic scattering,
instead of the inelastic scattering of the longitudinal
case. This affects the evaluation of $\cL^*=\cL$
on quadratic polynomials:
\beqn
\cL \bx&=& 0\\
\cL\bp&=&2\pi\int \delta(\frac{|\bp|^2-|\bq|^2}{2})
\int\Phi(w,\bp-\bq)dw(\bp-\bq)d\bq=0\\
\cL |\bx|^2&=&0\\
\cL \bx\cdot\bp&=&
0\\
\cL |\bp|^2&=&-2\pi\int \delta(\frac{|\bp|^2-|\bq|^2}{2})
\int\Phi(w,\bp-\bq)dw
\lt[ |\bp|^2-|\bq|^2\rt] d\bq =0.
\eeqn
The same results hold for the diffusion operator
$\cL$ with the diffusion matrix (\ref{D2}).

\subsection{Rate of dispersion}
The variance identity still holds
Hence the second derivative of $V_x$ becomes
\beq
\frac{\partial ^2}{\partial z^2} V_x
&=&4(H-\Hbar)+\frac{2(2-d\sigma)g}{\sigma+1}\int
\rho^{\sigma+1}d\bx
\label{trans-var1}\\
\label{trans-var2}
 &=&2d\sigma
(H-\Hbar)+(2-d\sigma)V_p.
\eeq

The main difference from the longitudinal case
is that the Hamiltonian is invariant
\beqn
\partz H=0.
\eeqn

By the same argument as in the longitudinal case
we have the following analogous estimates.
\begin{prop} 
If $d\sigma=2$ or $g=0$, then
\beq
\label{exact2}
V_x(z)=V_x(0)+2V_{xp}(0)z+2(H(0)-\Hbar)z^2,\quad
z\in [0, z_*).
\eeq 

\end{prop}
\begin{prop}
\label{prop6}
The following
estimates hold 
for $z\in [0, z_*)$:
\beqn
V_x(z)&\geq& V_x(0)+2V_{xp}(0)z+2(H-\Hbar)z^2,\quad
\hbox{for}\,\,g(2- d\sigma)\geq 0\\
V_x(z)&\leq& V_x(0)+2V_{xp}(0)z+2( H-\Hbar)z^2,\quad
\hbox{for}\,\,g(2-d\sigma)\leq 0
\eeqn
and 
\beqn
V_x(z)&\geq& V_x(0)+2V_{xp}(0)z+d\sigma(
H-\Hbar)z^2,\quad
\hbox{for}\,\,2\geq d\sigma\\
V_x(z)&\leq& V_x(0)+2V_{xp}(0)z+d\sigma(
H-\Hbar)z^2,\quad
\hbox{for}\,\,2\leq d\sigma.
\eeqn

\end{prop}

Therefore,
\begin{cor}
\label{cor2}
Assume $g< 0$ (hence $H\geq 0$). Then for
$ z\in [0, z_*)$
\beq\label{defocus2}
V_x(0)+2V_{xp}(0)z+(d\sigma\wedge 2) [H-\Hbar]z^2 \leq
V_x(z)\leq V_x(0)+2V_{xp}(0)z+(d\sigma\vee 2)
[H-\Hbar]z^2.
\eeq

\end{cor}
From these estimates we see that a ballistic kind of
motion takes place in the transverse case.

\subsection{Singularity}
By finding the zeros of the upper bounds in
Proposition~\ref{prop6} we can derive the conditions for
singularity formation. As in the longitudinal case, for $g\leq 0$,
(\ref{exact2}) and the left side of (\ref{defocus2})
are always positive.
\begin{prop} 
For $g>0, d\sigma \geq 2$, 
the local solution of the nonlinear kinetic eq. (\ref{qtg})
with the transverse randomness 
develops singularity in a finite time $z_*<\infty$
under either of the following conditions
\beq
\label{sing2.1}
 H<\Hbar
\eeq
or 
\beq
\label{sing2.2}
H\geq \Hbar, \quad V_{xp}(0)<0, \quad 
H\leq \Hbar  +|V_{xp}(0)|^2/(d\sigma V_x(0)).
\eeq
\end{prop}
\begin{remark}
It is easy to see that the singularity time $z_*$
is bounded from above by
\beq
\label{51}
\zo=\frac{-V_{xp}(0)+\sqrt{|V_{xp}(0)|^2-d\sigma V_x(0)(
 H- \Hbar)}}{d\sigma( H-\Hbar)}
\eeq
under condition (\ref{sing2.1}) and
by
\[
\zo=\frac{-V_{xp}(0)-\sqrt{|V_{xp}(0)|^2-d\sigma V_x(0)(
H-\Hbar)}}{d\sigma( H-\Hbar)}
\]
under condition (\ref{sing2.2}).
\end{remark}
The preceding singularity conditions are identical
to those for the homogeneous (super)-critical NLS
equation \cite{SS}. This is already suggested
by  the previous result \cite{nls} where
(\ref{sing2.1}) is shown to be the instability
condition for the diffusion approximation of 
the kinetic equation (\ref{qtg}) with the kernel
(\ref{K2}).

With more stringent conditions one can demonstrate
a stronger sense of blow-up. Let us state
the result.
\begin{prop}
Suppose 
$d\sigma>2, g> 0$. Then under the
conditions
\[H<\Hbar,
\]
$V_x'$ as well as $V_p$ blow up at a finite time
$z_*<z_0$.
\end{prop}

Let us sketch the argument below.
From (\ref{trans-var2}) it follows
that
\beq
\label{q2.2}
\frac{\partial ^2}{\partial z^2} V_x
&\leq &(2-d\sigma)V_p<0,\quad z<z_*.
\eeq
Since $V_x'(z)$ is a decreasing function and
becomes negative after a finite time $\zbar$
when $V_x(z)$ achieves its maximum $\bar{V}_x$.

Again for $A(z)=-V'_x(z)>0, z>\zbar$ we have from
(\ref{q2.2}) and (\ref{4.4}) the
differential inequality
\[
\partz A\geq CA^2,\quad
C=\frac{d\sigma-2}{4{V}_x(\zbar)}>0,\quad
z>\zbar.
\]

This is a plausible mechanism for the finite-time
blow-up
of the kinetic energy  and the potential energy.

\commentout{*******
\begin{remark}
The singularity formations condition 
for $g\leq 0$ based on 
Corollary~\ref{cor2}, namely
\beq
\label{sing3.2}
V_{xp}(0)<0, \quad 
H\leq \Hbar  +|V_{xp}(0)|^2/((d\sigma \vee 2)V_x(0)).
\eeq
or
\[
H<\Hbar
\]
can never be satisfied.
For instance, condition (\ref{sing3.2}) implies
that
\[
\frac{d\sigma\vee 2}{2}V_p(0)V_x(0)
-\frac{(d\sigma \vee 2)g}{\sigma+1}
\int \rho^{\sigma+1}d\bx \leq
V_{xp}=S_{xp}-\xbar\cdot\pbar
\]
which violates the Cauchy-Schwartz inequality.
\end{remark}
*************}

\commentout{***************************
************************************
Again, since  the variance identity (\ref{exact2}) is exact
for the critical nonlinearity, we can easily show
the following
\begin{prop}
Assume $d\sigma=2$. The solution of (\ref{qtg})
develops singularity at the finite time $z_*$ if
and only if either (\ref{sing2.1}) or (\ref{sing2.2})
holds.

If condition  (\ref{sing2.1}) holds then
\[
z_*=
\frac{-V_{xp}(0)+\sqrt{|V_{xp}(0)|^2-2 V_x(0)(
H- \Hbar)}}{2(H-\Hbar)};
\]
if condition (\ref{sing2.2}) holds, then
\[
z_*=
\frac{-V_{xp}(0)-\sqrt{|V_{xp}(0)|^2-2 V_x(0)(
H- \Hbar)}}{2(H-\Hbar)}.
\]
In general, the wave collapses
in the following way as $z\to z_*$
\beq
\label{rate1}
V_x(z) \sim (z_*-z), \quad z< z_*;
\eeq
except when 
\[
z_*=\frac{V_{xp}(0)}{2(\Hbar-H)}
\]
the rate of collapse becomes
\[
V_x(z) \sim (z_*-z)^2, \quad z< z_*.
\]
\end{prop}
Once again we see that the random scattering
has slowed down the rate of collapse from
the quadratic rate (\ref{rate2}) to
 the generic one of the linear (\ref{rate}), (\ref{rate1}).
 *******************************************}

\section{Fokker-Planck equation with dissipation}
In this section we will apply the same  analysis
to the Fokker-Planck equation (\ref{qtg}) with,
instead, 
the  Caldeira-Leggett operator \cite{CL}
\beq
\label{cl}
\cL W=D\Delta_\bp W+\lamb\gradp [\cdot\bp W],\quad
D, \lamb >0
\eeq
which is, perhaps, the simplest example of the 
phase-space models for open, dissipative
systems \cite{Dio}, \cite{AM}. As remarked before
in this context $z$ is the physical time, $\bx$
the physical coordinates and $\bp$ the momentum in $\IR^3$.
If the open quantum system is the electron motion in a semi-conductor
then the nonlinearity arises from the electrostatic
Hartree potential self-induced
by the density $\rho$ with self-defocusing
nonlinearity. Since operator (\ref{cl}) is not
self-adjoint (w.r.t. the Lebesgue measure)
it is outside the scope of the NLS
eq. with a real-valued random
potential (\ref{nls}). The method used here can be applied to
such a problem. However, in order to compare
how different decoherence mechanisms affect wave spreading
and collapse we will still consider the same power-law
nonlinearity as in the previous cases.

The dissipative
effect is of frictional nature and the Fokker-Planck
equation with (\ref{cl}) still preserves
the total mass
\[
\partz N=0.
\] 
As before we set $N=1$.

We re-calculate the  action of $\cL^*$, the adjoint
operator, on the quadratic polynomials:

\beqn
\cL^* \bx&=& 0\\
\cL^*\bp&=&-\lamb\bp\\
\cL^* |\bx|^2&=&0\\
\cL^* \bx\cdot\bp&=&-\lamb\bx\cdot\bp\\
\cL^* |\bp|^2&=&2dD-2\lamb |\bp|^2.
\eeqn
\subsection{Rate of Spread}
Using these results and integrating by parts
we get
\beqn
\partz \xbar&=&\pbar\\
\partz \pbar&=&-\lamb\pbar
\eeqn
which again constitute an exactly solvable system:
\beq
\label{6.2}
\pbar(z)&=&e^{-\lamb z} \pbar(0)\\
\xbar(z)&=&\xbar(0)+\frac{\pbar(0)}{\lamb}(1-e^{-\lamb
z}).\label{6.2'}
\eeq
The Hamiltonian $\Hbar$, defined by (\ref{Hbar}), of the
mean motion is then given by
\[
\Hbar(z)=\Hbar(0)e^{-2\lamb z}.
\]

Also, 
\beqn
\partz V_x
&=&2 V_{xp}\\
\partz V_{xp}
&=&d\sigma (H-\Hbar)+(1-\frac{d\sigma}{2})V_p-\lamb
V_{xp}\\ &=&2(H-\Hbar)+\frac{(2-d\sigma) g}{\sigma+1}
\int \rho^{\sigma+1}d\bx -\lamb V_{xp}.
\eeqn
\commentout{
or, alternatively,
\beqn
\partz S_{xp}
&=&2H+\frac{(2-d\sigma)g}{\sigma+1}\int
\rho^{\sigma+1}d\bx-\lamb S_{xp}
\eeqn
}
From these
we obtain
the upper bound
\beq
\label{6.10}
V_{xp}(z)&\leq&V_{xp}(0)+
d\sigma e^{-\lamb z}\int_0^z e^{s\lamb}
(H(s)-\Hbar(s))
ds,\quad\hbox{for}\,\,d\sigma \geq 2\\
V_{xp}(z)&\geq&V_{xp}(0)+
d\sigma e^{-\lamb z}\int_0^z e^{s\lamb}
(H(s)-\Hbar(s))
ds,\quad\hbox{for}\,\,d\sigma \leq 2
\eeq
 and
the bounds
\beq
\label{6.11}
V_{xp}(z)&\leq&V_{xp}(0)+
2e^{-\lamb z}\int_0^z e^{s\lamb}
(H(s)-\Hbar(s))
ds,\quad\hbox{for}\,\,g(2-d\sigma)\leq 0\\
V_{xp}(z)&\geq&V_{xp}(0)+
2e^{-\lamb z}\int_0^z e^{s\lamb}
(H(s)-\Hbar(s))
ds,\quad\hbox{for}\,\,g(2-d\sigma)\geq 0.
\eeq

\commentout{**********
 in the sub-critical case ($d\sigma
<2$), we have the lower bound
\beqn
\partz S_{xp}&\geq &2H
-\lamb S_{xp}\\
S_{xp}(z)&\geq &S_{xp}(0)+
d\sigma e^{-\lamb z}\int_0^z e^{s\lamb}
H(s)
ds.
\eeqn
*********}
 On the
other hand, the rate of change of the Hamiltonian is
given by
\beqn
\partz [H(z)-\Hbar(z)]&=& dD-\lamb V_p\\
&=& dD-2\lamb (H-\Hbar)-\frac{2\lamb g}{\sigma+1}
\int \rho^{\sigma+1}d\bx.
\eeqn
For self-focusing nonlinearity $g\geq  0$  the
upper bound on
$H$ follows
\beq
\label{hup}
H(z)-\Hbar(z)&\leq& H(0)-\Hbar(0)+ (1-e^{-2\lamb
z})\frac{dD}{2\lamb}
\eeq
whereas for the self-defocusing nonlinearity $g\leq 0$
we have both the lower as well the upper bounds
\[
H(0)-\Hbar(0)+ (1-e^{-2\lamb z})\frac{dD}{2\lamb}\leq
H(z)-\Hbar(z)\leq H(0)-\Hbar(0)+dD z.
\]
\commentout{******
for self-defocusing nonlinearity $g\leq 0$  the lower
bound follows
\beq
\label{hlow}
H(z)&\geq & H(0)+ (1-e^{-2\lamb z})\frac{dD}{2\lamb}
\eeq
****}
Therefore for $g\geq 0$ we have the upper  bounds 
\[
V_{xp}(z)\leq V_{xp}(0)+d\sigma
\lt[\frac{H(0)-\Hbar(0)}{\lamb}(1-e^{-\lamb z})
+(1-e^{-\lamb z})^2\frac{dD}{2\lamb^2}
\rt]
\]
whereas for $g\leq 0,$ we have the estimates
\beqn
V_{xp}(z)&\geq& V_{xp}(0)+(d\sigma\wedge 2)
\lt[\frac{H(0)-\Hbar(0)}{\lamb}(1-e^{-\lamb z})
+(1-e^{-\lamb z})^2\frac{dD}{2\lamb^2}
\rt]\\
V_{xp}(z)&\leq& V_{xp}(0)+(d\sigma\vee 2)
\lt[\frac{H(0)-\Hbar(0)}{\lamb}-\frac{dD}{\lamb^2}\rt](1-e^{-\lamb
z})+\frac{d^2\sigma D}{\lamb} z.
\eeqn
Therefore,
\begin{prop}
\label{prop8}
For $g\geq 0, z\in [0, z_*)$ 
\beq
\label{64}
V_x(z)&\leq&
V_x(0)+2V_{xp}(0)z+\frac{d\sigma}{\lamb}(2H(0)-2\Hbar(0)+\frac{dD}{\lamb})
z\\
&&-\frac{2d\sigma}{\lamb^2}\lt[H(0)-\Hbar(0)+\frac{dD}{\lamb}
\rt](1-e^{-\lamb z})
+\frac{d^2 \sigma D}{2\lamb^3}(1-e^{-2\lamb
z})\nn;
\eeq
whereas for $g\leq 0, z\in [0, z_*)$
\beqn
\label{6.6}
V_x(z)&\geq&
V_x(0)+2V_{xp}(0)z+\frac{d\sigma\wedge
2}{\lamb}(2H(0)-2\Hbar(0)+\frac{dD}{\lamb}) z\\
&&-\frac{2(d\sigma\wedge
2)}{\lamb^2}\lt[H(0)-\Hbar(0)+\frac{dD}{\lamb}
\rt](1-e^{-\lamb z})
+\frac{(d \sigma \wedge 2)dD}{2\lamb^3}(1-e^{-2\lamb
z})\\
V_x(z)&\leq&V_x(0)+2V_{xp}(0)z
+2(d\sigma\vee 2)\lt[\frac{H(0)}{\lamb}
-\frac{dD}{\lamb^2}\rt]
\lt[z+\frac{1}{\lamb}(e^{-\lamb z}-1)\rt]
+\frac{d^2\sigma D}{\lamb}z^2.
\eeqn

\end{prop}

\commentout{******
On the other hand, using (\ref{6.3}) and (\ref{decay})
we obtain that
\beqn
-2\int^z_* \xbar\cdot\pbar(s)
ds \geq -\frac{2}{\alpha}\Hbar(0)+|\xbar(0)|^2.
\eeqn
Now for 
$g\leq 0$ in the (sub)-critical case $d\sigma\leq 2$
the inequality in (\ref{6.6}) is reversed. Hence we
have the following
\begin{prop}
If $d\sigma \leq  2, g\leq 0$ then
for any smooth solution defined
on the interval $z\in [0, Z]$ 
\beqn
V_x(z)&\geq&
V_x(0)+\frac{d\sigma}{\lamb}(2H(0)+\frac{dD}{\lamb}) z
-\frac{2d\sigma}{\lamb^2}\lt[H(0)+\frac{dD}{\lamb}
\rt](1-e^{-\lamb z})\\
&&
+\frac{d^2 \sigma D}{2\lamb^3}(1-e^{-2\lamb
z})-\frac{2}{\alpha}\Hbar(0)+|\xbar(0)|^2,\quad
z\in [0, Z]
\eeqn
\end{prop}
*******}

\subsection{Singularity}
The singularity conditions for the self-focusing
case  are determined from that
the right side of (\ref{64}) has a positive root.
\begin{prop}
\label{sing-prop3}
For $d\sigma \geq 2, g> 0$, 
the local solution of the Fokker-Planck eq. (\ref{qtg}) 
with (\ref{cl})
develops finite-time singularity if 
\beq
\label{sing6}
V_{xp}(0)+\frac{d\sigma}{\lamb}(H(0)-\Hbar(0)+\frac{dD}{2\lamb})
<0.
\eeq
\end{prop}
The condition (\ref{sing6}) is better suited for
large $\lamb$ and 
can be much improved for small $\lamb$. As $\lamb\to 0$
this case should approach to that of
the longitudinal case with
$\bD$ given by (\ref{D1}) being a scalar.

Let us sketch an explicit
blow-up mechanism in the super-critical case 
under different conditions.

\begin{prop}
Suppose $d\sigma >2, g>0$. Then 
under the conditions
\beq
\label{70}
V_{xp}(0)\leq 0,\quad
H(0)-\Hbar(0)+\frac{dD}{2\lamb}
<0
\eeq
and
\beq
\label{72}
V_x(0)< \frac{d\sigma -2}{2\lamb }
\eeq
$V_x'$ as well as $V_p$ blow up at a finite
time $z_*$ where
\[
z_*<\frac{1}{\lamb}\ln{\frac{d\sigma -2}{d\sigma-2
-2\lamb V_x(0)}}.
\]
\end{prop}
 
We present the argument here.  Rewriting some of
the preceding calculations, we have
\[
\partz V_x=2V_{xp}
\]
and
\beq
\label{69}
\partz V_{xp}=d\sigma (H-\Hbar)+(1-\frac{d\sigma}{2})V_p
-\lamb V_{xp}.
\eeq

Condition (\ref{70}) and (\ref{hup})
imply that
\[
H(z)-\Hbar(z)\leq -e^{-2\lamb z}\frac{dD}{2\lamb}.
\]
Using this in (\ref{69}) we obtain
\[
\partz (e^{\lamb z} V_{xp}(z))\leq
-e^{-\lamb z}\frac{d^2\sigma D}{2\lamb}
+e^{\lamb z}(1-\frac{d\sigma}{2})V_p
\]
which together with (\ref{70}) implies
$V_{xp}(z)<0, z\leq z_*$ and $V_x(z)$
is decreasing. Using the inequality
\[
V_p\geq \frac{V_{xp}^2}{V_x(0)}
\]
in (\ref{69}) we obtain again the quadratic
growth estimate
\[
\partz A\geq \frac{d\sigma -2}{2V_x(0)}
A^2-\lamb A,\quad z< z_*
\]
for $A=-V_{xp}>0$, or equivalently
\[
\partz B\geq \frac{d\sigma -2}{2V_x(0)}
e^{-\lamb z}B^2,\quad z<z_*
\]
for
\[
B=e^{\lamb z} A.
\]
From this it follows that
\[
B(z)\geq \frac{B(0)}{1
-\frac{d\sigma -2}{2\lamb V_x(0)}(1-e^{-\lamb z})}
\]
the right-hand side of which blows up at a finite time
if and only if
(\ref{72}) holds.

The finite-time blow-up of $B$ would also lead
to the finite-time blow-up of $A$ and hence of
$V_p$.

\section{Conclusion}

We have derived several kinetic equations 
meant to model the large scale, low Fresnel number
behavior of the NLS equation with a rapidly
fluctuating random potential based on the rigorous
theory
\cite{rad-arma}, \cite{saq} for the linear
Schr\"odinger waves in the same situation.
The low Fresnel number waves interacting  with
a rapidly oscillating potential
give rise to, in the scaling (\ref{scale}),  a self-averaging
limit of a deterministic 
kinetic equation with a scattering
operator. 

We analyze these kinetic equations
in order to shed light on two problems: the rate of
dispersion and the singularity formation.
Our main assumptions are the existence of
local solutions of the kinetic models in the space
$\cS$, that the local solutions satisfy the Hamiltonian
laws (\ref{h1}), (\ref{h2})
and  the variance identity.

The scattering  operator in the kinetic equations
may be a Boltzmann-type operator or
a Fokker-Planck operator, depending
on whether the random potential
is fully resolved by the wave or not. 
This, however, does not affect
our results for either problem.

What is more important  for our investigation is
the  {\em structure} of the scattering kernel in
the Boltzmann operator or the diffusion matrix
of the Fokker-Planck operator. 
There are two types of  structures:
the longitudinal regime when the longitudinal
scale of the random potential is dominant
and the transverse regime when
the transverse scale of the random potential
is dominant. The third kind
of kinetic equation that we have analyzed
is based on the Fokker-Planck equation
with the longitudinal type of diffusion matrix
plus a frictional term. This model is motivated
by the Caldeira-Leggett operator for  open quantum
dynamics.

For the problem of dispersion, we have shown
that the kinetic equations of the longitudinal type
produce the cubic ($z^3$) law, that the kinetic equations
of the transverse type produce the quadratic
($z^2$) law  and that the Fokker-Planck
equation with friction produces
the linear ($z$) law  for
the variance $V_x(z)$ for $z\in [0, z_*)$.

For the problem of singularity, we have shown
by analyzing the variance identity
that the singularity and blow-up conditions
in the transverse
case remain the same as those for the homogeneous
NLS equation with critical or supercritical self-focusing
nonlinearity, that the singularity and
blow-up conditions have changed in
the longitudinal case and in the frictional case.

Finally let us make a slight extension
of our results by considering
the nonlinear kinetic equation
of the more general form
\beq
\label{qtg2}
\partz W(z,\bx,\bp)+\bp\cdot\nabla_\bx W
(z,\bx,\bp)+\cU_\gamma
W(z,\bx,\bp)=\cL W(z,\bx,\bp)
\eeq
with $\gamma>0$ small, but positive and the mixed-state Wigner
distributions (\ref{mixed}) as initial data.
 Eq. (\ref{qtg}) is
formally   the  geometrical optics limit of eq.
(\ref{qtg2}). 
Eq. (\ref{qtg2})  retains the diffraction effect of the
nonlinear potential term that is absent
in eq. (\ref{qtg}) with $\gamma=0$. Even though
eq. (\ref{qtg2})  does not  preserve
positivity, 
it can be shown to preserve the mixed state structure
(\ref{mixed}) and therefore the marginal positivity
(\ref{pos1}), (\ref{pos2})
as well as the inequalities
(\ref{cs2}), (\ref{cs})
in both the longitudinal and
transverse cases. Since the calculations
(\ref{ga1})-(\ref{ga2}) are independent
of $\gamma\geq 0$, all the results in
Section~4 and 5 hold
for eq. (\ref{qtg2}) with $\gamma>0$ as well.
However, because the Caldeira-Leggett
operator (\ref{cl}) is in the Lindblad form only in
the classical limit \cite{Dio}, therefore in
order to maintain the solution in the 
the mixed-state form (\ref{mixed}) we 
 need to set $\gamma=0$ in this particular case.

\end{document}